\documentclass{article}
\usepackage{times}
\usepackage{amsfonts}
\begin{document}
\title{Nonabelian Poisson Manifolds from D--Branes}
\author{Jos\'e M. Isidro\\
Instituto de F\'{\i}sica Corpuscular (CSIC--UVEG)\\
Apartado de Correos 22085, Valencia 46071, Spain\\
and\\
Max-Planck-Institut f\"ur Gravitationsphysik, 
\\Albert-Einstein-Institut, \\
D-14476 Golm, Germany\\
{\tt jmisidro@ific.uv.es}}

\hyphenation{di-men-sion}
\hyphenation{di-men-sion-al}
\maketitle

\begin{abstract}

Superimposed D--branes have matrix--valued functions as their transverse coordinates, since the latter take values in the Lie algebra of the gauge group inside the stack of coincident branes. This leads to considering a classical dynamics where the multiplication law for coordinates and/or momenta, being given by matrix multiplication, is nonabelian. Quantisation further introduces noncommutativity as a deformation in powers of Planck's constant $\hbar$. Given an arbitrary simple Lie algebra $\mathfrak{g}$ and an arbitrary Poisson manifold ${\cal M}$, both finite--dimensional, we define a corresponding $C^{\star}$--algebra that can be regarded as a {\it nonabelian Poisson manifold}. The latter provides a natural framework for a matrix--valued classical dynamics.

\end{abstract}

\tableofcontents

\section{Introduction}\label{ramalloguarrolavatelospies}

Let ${\cal M}$ denote the classical phase space of a classical dynamics with a finite number of degrees of freedom governed by a time--independent Hamiltonian function $H$. As such ${\cal M}$ will be a finite--dimensional Poisson manifold, with $C^{\infty}({\cal M})$ as its algebra of smooth functions. For technical reasons we will be mostly interested in the case when ${\cal M}$ is compact. Upon quantisation, ${\cal H}$ will denote the Hilbert space of quantum states. Classical functions on phase space $f\in C^{\infty}({\cal M})$ become quantum operators $F$ on Hilbert space. This we denote as $F\in{\cal O}({\cal H})$, where ${\cal O}({\cal H})$ stands for the algebra of observables. When ${\cal M}$ is compact, ${\cal H}$ is finite--dimensional, and so is ${\cal O}({\cal H})$ too. We use lowercase letters $f$ for classical functions and uppercase letters $F$ for their quantum counterparts; the only exception to this rule is the Hamiltonian, denoted $H$ both as a classical function and as a quantum operator. All functions and all operators will be time--independent.

The algebra ${\cal C}^{\infty}({\cal M})$ supports classical Poisson brackets (CPB), {\it i.e.}, an antisymmetric, bilinear map
\begin{equation}
\left\{\cdot\,,\cdot\right\}_{\rm CPB}\colon {\cal C}^{\infty}({\cal M})\times {\cal C}^{\infty}({\cal M})\longrightarrow {\cal C}^{\infty}({\cal M})
\label{labastidaquetepartaunrayo}
\end{equation}
satisfying the Jacobi identity and the Leibniz derivation rule.
Upon quantisation, the algebra of functions $C^{\infty}({\cal M})$ on classical phase space gets replaced by the algebra of operators ${\cal O}({\cal H})$ on Hilbert space. 
The quantum Poisson bracket $\left[\cdot\,,\cdot\right]_{\rm QPB}$ is an antisymmetric, bilinear map
\begin{equation}
\left[\cdot\,,\cdot\right]_{\rm QPB}\colon {\cal O}({\cal H})\times {\cal O}({\cal H})\longrightarrow {\cal O}({\cal H})
\label{labastidacabronquetepartaunrayo}
\end{equation}
also satisfying the Jacobi identity and the Leibniz derivation rule.

In this letter we extend the algebra of functions on a Poisson manifold to become a Poisson algebra that also encodes
degrees of freedom associated with a simple Lie algebra. Based on a remark made in ref. \cite{WITTEN}, we observe that the M(atrix) theory action \cite{BFSS} provides us with a dynamical system whose classical phase space requires the notion of Lie--algebra valued coordinates and momenta. We use this approach to define a $C^{\star}$--algebra that can be regarded as a nonabelian Poisson manifold. We distinguish between the terms {\it nonabelian}\/ and {\it noncommutative}, reserving the former for multiplication laws such as that for matrices, and the latter for multiplication laws such as $\star$--products involving $\hbar$--deformations. Finally we extend our technique to construct more general classes of $C^{\star}$--algebras, based on manifolds taken to lie orthogonally to a stack of coincident branes.

Matrix quantum mechanics has been studied recently in ref. \cite{LIZZI} in connection with noncommutative field theory. The deformation quantisation for affine Poisson varieties has been analysed in ref. \cite{LLEDO}. Along an apparently unrelated line, it has been argued that the internal dynamics of a baryon, as a bound state of QCD--strings and quarks, may be captured by a theory of matrix coordinates \cite{FATOLLAHI}. We should stress that the deformation quantisation of Poisson manifolds \cite{KONTSEVICH}, 
although related with our subject, is not addressed here.

\section{D--branes and Poisson brackets}\label{casposoramallo}

\subsection{Classical Poisson brackets in the absence of Lie--algebra symmetries}\label{kabronbarbon}

The superposition of $n$ parallel, identical D$p$--branes produces a $u(n)$ gauge theory on their common $(p+1)$--dimensional worldvolume \cite{WITTEN}.  Although this theory is supersymmetric, we will concentrate throughout on its bosonic sector. Now ${u}(n)={u}(1)\times {su}(n)$ is not simple, but separating out the centre--of--mass motion we are left with the simple Lie algebra ${su}(n)$. Let $A_{\mu}$ be an ${su}(n)$--valued gauge field on the D--brane stack, and let us separate its components into longitudinal and transverse parts to the D--branes, $A_{\mu}=(A_l, A_t)$. Longitudinal components $A_l$ are adjoint--valued ${su}(n)$ matrices, {\it i.e.}, Yang--Mills gauge fields. Transverse components $A_t$ describe D--brane fluctuations that are orthogonal to the D--branes. They are thus identified with transverse coordinates, so they are more properly denoted $x_l$ instead of $A_l$. (To conform with our convention we reserve the notation $X_l$ for the quantum operator corresponding to the classical function $x_l$). Modulo numerical factors, the bosonic part of super Yang--Mills theory dimensionally reduced to $p+1$ dimensions is
\begin{equation}
S_{\rm YM}^{(p+1)}=\int{\rm d}^{p+1}\xi\,{\rm tr}\,({\cal F}_{ll'}^2+2{\cal F}^2_{lt}+{\cal F}^2_{tt'}),
\label{labastidachupamelapollamarikondemierda}
\end{equation}
where $l,l'$ ($t,t'$) are longitudinal (transverse) indices. D--boundary conditions remove all derivatives in the $t$ directions, and (again up to numerical factors) eqn. (\ref{labastidachupamelapollamarikondemierda}) becomes
\begin{equation}
S_{\rm YM}^{(p+1)}=\int{\rm d}^{p+1}\xi\,{\rm tr}\,{\cal F}_{ll'}^2 - \int{\rm d}^{p+1}\xi\,{\rm tr}\,\left(\frac{1}{2} (D_lx^t)^2-\frac{1}{4}[x^t,x^{t'}]^2
\right),
\label{cesargomezquetefollenkabrondemierda}
\end{equation}
where $D_lx^t=\partial_lx^t+{\rm i}[A_l, x^t]$ is the longitudinal, gauge--covariant derivative of transverse coordinates $x^t$. The appearance of matrix--valued coordinate functions can be motivated in the relation of D$p$--branes to Chan--Paton factors via T--duality. When $p=0$ we have the important case of the M(atrix) model of M--theory in the light--cone gauge \cite{BFSS}, where the limit $n\to\infty$ is taken.

We are interested in the {\it transverse}\/ coordinates to the D--brane. They are described by the terms in eqn. (\ref{cesargomezquetefollenkabrondemierda}) that are not the Yang--Mills field strength,
\begin{equation}
S=- \frac{1}{2}\int{\rm d}^{p+1}\xi\,{\rm tr}\,\left(\frac{1}{2} (D_lx^j)(D_lx^j)-\frac{1}{2}\sum_{i\neq j}[x^i,x^{j}]^2
\right),
\label{luisibanezquetefollenkabrondemierda}
\end{equation}
where $l=0, 1,, \ldots p$ runs over the longitudinal coordinates to the D--brane, and $i,j=p+1, \ldots, d$ run over the transverse coordinates to the D--brane. The latter is embedded within $d$--dimensional spacetime $\mathbb{R}^d$, with a metric $(-,+,+,\ldots, +)$. In M--theory $d=11$, for strings we have $d=10$. The $\xi^l$ are the longitudinal worldvolume coordinates on the D--branes that the transverse functions $x^j=x^j(\xi)$ depend on. Being matrices, the $x^j(\xi)$ are Lie--algebra valued,
\begin{equation}
x^j(\xi)=\sum_{a=1}^{n^2-1}x_a^j(\xi)T_a,
\label{barbonquetefollenhijoputa}
\end{equation}
where the $T^a$ generate $su(n)$ in the adjoint representation. To the Lagrangian density of eqn. (\ref{luisibanezquetefollenkabrondemierda}),
\begin{equation}
{\rm l}=-\frac{1}{2}{\rm tr}\,(D_lx^jD_lx^j)+\frac{1}{4}{\rm tr}\,\sum_{i\neq j}[x^i,x^j]^2,
\label{mekagoentodoslosdelaautonoma}
\end{equation}
there corresponds a Hamiltonian density
$$
{\rm h}=\frac{1}{2}{\rm tr}\,(p^jp^j)+\frac{1}{2}{\rm tr}\,(\partial_sx^j\partial_sx^j)+{\rm i}\,{\rm tr}\,(p^j[A_0,x^j])
$$
\begin{equation}
+{\rm i}\,{\rm tr}\,(\partial_sx^j[A_s,x^j])-\frac{1}{2}{\rm tr}\,[A_s, x^j]^2-\frac{1}{4}\sum_{i\neq j}{\rm tr}\,[x^i,x^j]^2,
\label{labastidamekagoentuputakara}
\end{equation}
where the subindex $s$ stands for the spacelike, longitudinal coordinates $\xi^1, \ldots, \xi^p$, and the (adjoint--valued) $p^j$ are the canonical momenta conjugate to the $x^j$. The (equal--time) CPB between coordinates and momenta for this field theory read
\begin{equation}
\left\{x^i_a(\xi^0,\xi), p^k_b(\xi^0, \xi')\right\}_{\rm CPB}=\delta_{ab}\delta^{ik}\delta^{(p}(\xi-\xi'),
\label{mekagoentuputakararamallo}
\end{equation}
where $\delta^{(p}(\xi-\xi')$ refers to the $p$ spacelike, longitudinal coordinates along the D$p$--brane. This delta function disappears when $p=0$, in which case the CPB (\ref{mekagoentuputakararamallo}) simplify from those of a field theory to those of a finite number of degrees of freedom,
\begin{equation}
\left\{x^i_a(\xi^0), p^k_b(\xi^0)\right\}_{\rm CPB}=\delta_{ab}\delta^{ik}.
\label{mekagoentuputakaralabastidademierda}
\end{equation}
In what follows we will set $p=0$ in the action (\ref{luisibanezquetefollenkabrondemierda}), so the corresponding CPB are given by eqn.  (\ref{mekagoentuputakaralabastidademierda}). Then classical phase space ${\cal M}$ has the $2d(n^2-1)$ Darboux coordinates $x^{i}_a$ and $p^{k}_b$: there are $d$ transverse coordinates to a D$0$--brane, all of which are $su(n)$--valued. Hereafter $C^{\infty}({\cal M})$ will denote the algebra of smooth functions in the variables $x^{i}_a$ and $p^{k}_b$.

\subsection{Classical Poisson brackets in the presence of Lie--algebra symmetries}\label{kabronbarbonmarikon}

The Hamiltonian density (\ref{labastidamekagoentuputakara}) is a Lie--algebra scalar. However it is natural to consider functions of the matrix variables
\begin{equation}
x^{i}:=x^{i}_aT_a,\qquad p^{k}:=p^{k}_aT_a,\qquad x^{i}_a, p^{k}_b\in C^{\infty}({\cal M})
\label{barbonketepartaunrayo}
\end{equation}
having higher transformation properties (under the Lie algebra) than those of a scalar. 
Having {\it matrix--valued coordinates and momenta}\/ as in eqn. (\ref{barbonketepartaunrayo}) requires extending our algebra of functions.

{}For generality let us consider an arbitrary simple, real, finite--dimensional, compact Lie algebra $\mathfrak{g}$; eventually we will set $\mathfrak{g}=su(n)$. (Orthogonal and symplectic gauge groups can be obtained by adding orientifolds to the stack of $n$ coincident branes as done, {\it e.g.}, in ref. \cite{THEISEN}). 
Now $\mathfrak{g}$ supports Lie brackets
\begin{equation}
[\cdot\,,\cdot]\colon \mathfrak{g}\times \mathfrak{g}\longrightarrow \mathfrak{g}
\label{ramallocabron}
\end{equation}
which, in the basis $T_a$, read
\begin{equation}
[T_a,T_b]=\omega_{ab}^{\;\; c}T_c.
\label{mierdaparavr}
\end{equation}
We will consider the $T_a$ in the adjoint representation. The universal enveloping algebra ${\cal U}(\mathfrak{g})$ contains the quadratic Casimir operator $k^{ab}T_aT_b=c_A{\bf 1}$. Here $k^{ab}$ is the inverse Killing metric. Since $\mathfrak{g}$ is compact 
we can henceforth assume that the basis $T_a$ is orthonormal, 
\begin{equation}
T_aT_a=c_A{\bf 1}. 
\label{mekagoentuputakarabarbondemierda}
\end{equation}

Let us consider the tensor product algebra
\begin{equation}
{\cal C}({\cal M}, \mathfrak{g}):=C^{\infty}({\cal M})\otimes {\cal U}(\mathfrak{g}).
\label{ramallochupameelcarallo}
\end{equation}
Now ${\cal C}({\cal M}, \mathfrak{g})$ will qualify as a Poisson algebra if we can endow it with some Poisson brackets
\begin{equation}
\{\cdot\,, \cdot\}_{{\cal C}(\cal M,\mathfrak{g})}\colon {\cal C}({\cal M}, \mathfrak{g})\times {\cal C}({\cal M}, \mathfrak{g})\longrightarrow {\cal C}({\cal M}, \mathfrak{g}).
\label{avzrmekago}
\end{equation}
For this it suffices to define the canonical bracket $\{x^{i}, p^k\}_{{\cal C}({\cal M}, \mathfrak{g})}$, as any other bracket then follows by requiring antisymmetry, bilinearity and the Leibniz derivation rule. Using eqn. (\ref{mekagoentuputakaralabastidademierda}) we set, in the adjoint representation for $\mathfrak{g}$,
\begin{equation}
\{x^{i},p^k\}_{{\cal C}({\cal M}, \mathfrak{g})}:=\{x^{i}_a, p^k_{b}\}_{\rm CPB}\,T_aT_b=\delta^{ik}\delta_{ab}\,T_aT_b=\delta^{ik}\,T_aT_a,
\label{putonramallo}
\end{equation}
{\it i.e.},
\begin{equation}
\{x^{i},p^k\}_{{\cal C}({\cal M}, \mathfrak{g})}=\delta^{ik}c_A{\bf 1}.
\label{ramallomekagoentuputakara}
\end{equation}
Then the Jacobi identity holds. This turns ${\cal C}({\cal M}, \mathfrak{g})$ into a Poisson algebra. The underlying $\mathfrak{g}$ appears through its quadratic Casimir eigenvalue $c_A$. In a representation $R$ other than the adjoint, $c_A$ is replaced with the corresponding quadratic Casimir eigenvalue $c_R$. The right--hand side of (\ref{ramallomekagoentuputakara}) is central within ${\cal C}({\cal M}, \mathfrak{g})$ as it should. For $\mathfrak{g}=su(n)$ we have $c_A=n$, and the fundamental brackets (\ref{ramallomekagoentuputakara}) read
\begin{equation}
\{x^{i},p^k\}_{{\cal C}({\cal M}, su(n))}=\delta^{ik}n{\bf 1}.
\label{ramallohijoputamekagoentuputakara}
\end{equation}
{}Finally the time evolution of an arbitrary $f\in {\cal C}({\cal M}, \mathfrak{g})$ is given by
\begin{equation}
\frac{{\rm d}f}{{\rm d}t}=\{f, H\}_{{\cal C}({\cal M}, \mathfrak{g})},
\label{kap}
\end{equation}
where $H$ is the Hamiltonian function (equal to the Hamiltonian density (\ref{labastidamekagoentuputakara}) because $p=0$).

\subsection{The quantum theory}\label{barbobchupamelapollakabron}

In order to quantise the dynamics of section \ref{kabronbarbonmarikon} let us first assume that we {\it turn off}\/ the Lie--algebra degrees of freedom. This is best achieved by separating all $n$ branes from each other, so no two of them remain coincident \cite{WITTEN}. Then $su(n)$ breaks into $n-1$ copies of $u(1)$. Effectively we are left with $n-1$ independent copies of $C^{\infty}({\cal M})$,
placed along the diagonal of an $(n^2-1)\times (n^2-1)$ matrix. Now $C^{\infty}({\cal M})$ can be quantised by standard methods to yield the algebra ${\cal O}({\cal H})$ of quantum observables on Hilbert space ${\cal H}$. After this operation we let all $n$ branes coincide again, and we consider the tensor product algebra
\begin{equation}
{\cal Q}({\cal H}, \mathfrak{g}):={\cal O}({\cal H})\otimes {\cal U}(\mathfrak{g}),
\label{ramalloeresunkasposdemierda}
\end{equation}
which is the quantum analogue of the classical algebra (\ref{ramallochupameelcarallo}).
Now ${\cal Q}({\cal H}, \mathfrak{g})$ will qualify as an algebra of quantum operators if we can endow it with quantum Poisson brackets
\begin{equation}
[\cdot\,,\cdot]_{{\cal Q}({\cal H}, \mathfrak{g})}\colon
{\cal Q}({\cal H}, \mathfrak{g})\times {\cal Q}({\cal H}, \mathfrak{g})\longrightarrow {\cal Q}({\cal H}, \mathfrak{g}).
\label{labastidamarikononquetepartaunrayo}
\end{equation}
{}For this it suffices to define $[X^{i},P^k]_{{\cal Q}({\cal H}, \mathfrak{g})}$ as the quantum analogue of eqn. (\ref{ramallomekagoentuputakara}),
\begin{equation}
[X^{i},P^k]_{{\cal Q}({\cal H}, \mathfrak{g})}={\rm i}\hbar\delta^{ik}c_A{\bf 1}.
\label{ramallomekagoentuputamama}
\end{equation}
Then ${\cal Q}({\cal H}, \mathfrak{g})$ qualifies as a Poisson algebra. That is, extending the brackets (\ref{ramallomekagoentuputamama}) to all ${\cal Q}({\cal H}, \mathfrak{g})$ by requiring linearity, antisymmetry and the Leibniz derivation rule automatically ensures that the Jacobi identity is satisfied. Finally the time evolution of an arbitrary $F\in {\cal Q}({\cal H}, \mathfrak{g})$ is given by the quantum counterpart of eqn. (\ref{kap}),
\begin{equation}
{\rm i}\hbar\frac{{\rm d}F}{{\rm d}t}=[F, H]_{{\cal Q}({\cal H}, \mathfrak{g})}.
\label{kappp}
\end{equation}

\section{The space orthogonal to the brane stack}\label{kabronramallo}

The $\mathfrak{g}$--valued classical coordinates $x^{i}$, $i=1, \ldots, d$, are orthogonal to the stack of $n$ coincident D$0$--branes. 
Further considering the corresponding momenta $p^k$, in section \ref{kabronbarbonmarikon} we have constructed an algebra $C({\cal M}, \mathfrak{g})$ of classical functions that defines a {\it nonabelian Poisson algebra}. Nonabelianity is a simple consequence of the matrix--valued character of the coordinates orthogonal to the stack of coincident branes. 

Analogous properties hold in the quantum case of section \ref{barbobchupamelapollakabron}: we have coordinate and momentum operators $X^{i}$ and $P^k$, and an algebra of quantum operators ${\cal Q}({\cal H}, \mathfrak{g})$ that defines a {\it nonabelian, noncommutative, Poisson algebra}. Nonabelianity means the same as in the classical case, while {\it noncommutativity}\/ refers to the presence of $\hbar\neq 0$.
Thus noncommutativity can be eliminated by passing to the classical limit $\hbar\to 0$, while nonabelianity remains even in the classical case.

We can take further advantage of the nonabelian property by simply disregarding their conjugate momenta $p^k$ and letting the $x^i$ cover a certain manifold ${\cal K}$. In string or M--theory one would usually require ${\cal K}$ to be compact. Depending on the amount of supersymmetry one wishes to preserve, typical examples for ${\cal K}$ could be a Riemann surface, a K3, or a Calabi--Yau manifold, among others. The compact manifold ${\cal K}$ leads to a nonabelian algebra defined as the tensor product
\begin{equation} 
{\cal C}({\cal K}, \mathfrak{g}):=C^{\infty}({\cal K})\otimes {\cal U}(\mathfrak{g})
\label{marikonramallochupameelcarallo}
\end{equation}
where $C^{\infty}({\cal K})$ is the algebra of smooth functions on ${\cal K}$. However, for as long as none of the $p^k$ are coordinates on  ${\cal K}$, the algebra ${\cal C}({\cal K}, \mathfrak{g})$ will not be noncommutative in the sense of $\hbar$--deformations. It will just be nonabelian.

The amount of conserved supersymmetry in (\ref{marikonramallochupameelcarallo}) is the complement of the one in which the branes are taken to wrap around ${\cal K}$. This situation is complementary to the one considered in ref. \cite{NOI}, when a noncommutative Riemann surface arose by wrapping the branes on a Riemann surface and turning on a background $B$--field \cite{CDS, SW}. Instead, here the D$p$--branes wrap a Minkowskian $\mathbb{R}^{p+1}$ in the absence of a Neveu--Schwarz $B$--field and are orthogonal to a certain compact ${\cal K}$. 

When do the algebras (\ref{ramallochupameelcarallo}), (\ref{ramalloeresunkasposdemierda}) and (\ref{marikonramallochupameelcarallo})
actually give rise to {\it manifolds}, eventually noncommutative manifolds? Whenever the corresponding algebra qualifies as a $C^{\star}$--algebra, it has an interpretation as a manifold \cite{NCG}, eventually noncommutative. Let us recall that a sufficient condition
for the algebra of continuous functions $C^0(V)$ on a topological space $V$ to be a $C^{\star}$--algebra is that $V$ be compact \cite{NCG}. On the other hand, the universal enveloping algebra ${\cal U}(\mathfrak{g})$ is a $C^{\star}$--algebra. Thus, requiring ${\cal K}$ in eqn. (\ref{marikonramallochupameelcarallo}) to be compact, we are assured that ${\cal C}({\cal K}, \mathfrak{g})$ qualifies as a $C^{\star}$--algebra, certainly nonabelian (because of matrix multiplication) although not yet noncommutative in the sense of $\hbar$--deformations. However, noncommutativity can be obtained if we require ${\cal M}$ in eqn. (\ref{ramallochupameelcarallo}) to be compact. Then ${\cal H}$ in eqn. (\ref{ramalloeresunkasposdemierda}) is finite--dimensional, and so is ${\cal O}({\cal H})$. This turns ${\cal Q}({\cal H}, \mathfrak{g})$ into a $C^{\star}$--algebra that is both nonabelian (because of the Lie algebra $\mathfrak{g}$) and noncommutative (because the quantum theory provides a deformation in powers of $\hbar$). In this case we may well call ${\cal Q}({\cal H}, \mathfrak{g})$ a {\it nonabelian, noncommutative Poisson manifold}.

\section{Discussion}\label{ramallotienesmuchacaspa}

In this letter we have presented a framework to describe the classical and quantum dynamics of Lie--algebra valued coordinates and momenta. The corresponding classical action is given by eqn. (\ref{luisibanezquetefollenkabrondemierda}) with $p=0$; it describes the {\it transverse}\/ excitations to a stack of $n$ coincident D$0$--branes.  This is interpreted in ref. \cite{WITTEN} as meaning that the coordinates orthogonal to the coincident branes, once quantised,  describe quantum fluctuations of the branes themselves. 

Given a simple Lie algebra $\mathfrak{g}$ and a compact Poisson manifold ${\cal M}$, both finite--dimensional, we have constructed the nonabelian $C^{\star}$--algebra ${\cal C}({\cal M}, \mathfrak{g})$. Nonabelianity arises from the multiplication of matrices as coordinates. It originates in the mere presence of a stack of $n$ parallel, coincident D--branes, inside which a (supersymmetric) $\mathfrak{g}$--valued Yang--Mills theory is defined. An observer sitting inside the brane stack cannot get around the fact that the coordinate functions orthogonal to the branes are Lie--algebra valued, hence nonabelian under multiplication. Further quantising this classical theory provides a deformation in powers of $\hbar$, of the sort usually termed {\it noncommutative}. This has led to the nonabelian, noncommutative $C^{\star}$--algebra ${\cal Q}({\cal H}, \mathfrak{g})$. Our construction can be extended to the case where one considers a compactificaction manifold ${\cal K}$ lying along the directions orthogonal to the stack of coincident branes. Then a $C^{\star}$--algebra ${\cal C}({\cal K},\mathfrak{g})$ can be defined that provides a nonabelian counterpart for the commutative algebra of functions $C^{\infty}({\cal K})$. 

In string theory, noncommutativity arises as the result of turning on a $B$--field \cite{CDS, SW}. Here noncommutativity 
is the result of quantising, {\it i.e.}, of deforming in powers of $\hbar$, those coordinates that are transverse to the branes.
Having $\hbar\neq 0$ or $\hbar=0$ {\it outside the branes}\/ plays the role of turning the $B$--field on and off {\it within the branes}. On the other hand, nonabelianity arises as a consequence of having Lie--algebra valued coordinates and momenta. The latter originate in the branes that lie orthogonally to the compactification manifold. Nonabelianity {\it outside the branes}\/ is turned on and off by having the branes coincide or by separating them. 

Two variations on our theme are worth considering. One is the case when classical phase space ${\cal M}$ is noncompact, so the Hilbert space ${\cal H}$ becomes infinite--dimensional. Then the algebra ${\cal B}({\cal H})$ of bounded operators on ${\cal H}$ is a $C^{\star}$--algebra \cite{NCG}, and a natural generalisation for ${\cal O}({\cal H})\otimes {\cal U}(\mathfrak{g})$ is ${\cal B}({\cal H})\otimes {\cal U}(\mathfrak{g})$. Another variation is to replace $C^{\infty}({\cal K})$ in eqn. (\ref{marikonramallochupameelcarallo}) with the $C^{\star}$--algebra $C^{\star}({\cal K})$ obtained by wrapping an independent set of $n'$ coincident D$p'$--branes on ${\cal K}$ and turning on a background $B$--field across the latter. Then the amount of conserved supersymmetry depends, among other things, on the relative orientation between the two stacks of coincident branes under consideration. It would be interesting to explore how the $B$--field across the second stack of $n'$ coincident D$p'$--branes influences the value of $\hbar$ on the first stack. We hope to report on these issues in the future.

\vskip1cm

{\bf Acknowledgements}

It is a great pleasure to thank J. de Azc\'arraga for encouragement and support. The author thanks S. Theisen and Max-Planck-Institut f\"ur Gravitationsphysik, Albert-Einstein-Institut for hospitality. This work has been partially supported by research grant BFM2002--03681 from Ministerio de Ciencia y Tecnolog\'{\i}a, by EU FEDER funds, by Fundaci\'on Marina Bueno and by Deutsche Forschungsgemeinschaft.

\end{document}